\documentclass[epjCONF]{svjour}
\usepackage{graphics}
\usepackage[varg]{txfonts} 
\usepackage[latin1]{inputenc}
\session-title{A Universe of Dwarf Galaxies: 2010, June 14-18, Lyon, France}
\begin{document}
\title{The Luminosity Function in Groups of Galaxies}
\author{R. Brent Tully\thanks{\email{tully@ifa.hawaii.edu}}}
\institute{Institute for Astronomy, University of Hawaii, Honolulu, Hawaii, USA}
\abstract{
With targeted imaging of groups in the local volume, the regions of collapse around bright galaxies can be clearly identified by the distribution of dwarfs and luminosity functions can be established to very faint levels.  In the case of the M81 Group there is completion to $M_R \sim -9$.  In all well studied cases, the faint end slopes are in the range $-1.35 < \alpha < -1.2$, much flatter than the slope for the bottom end of the halo mass spectrum anticipated by $\Lambda$CDM hierarchical clustering theory.  Small but significant variations are found with environment.  Interestingly, the populations of dwarf galaxies are roughly constant per unit halo mass.  With the numbers of dwarfs as an anchor point, evolved environments (dominated by early morphological types) have relatively fewer intermediate luminosity systems and at least one relatively more important galaxy at the core.  The variations with environment are consistent with a scenario of galaxy merging.  However it is questionable if the universal dearth of visible dwarf systems is a consequence of an astrophysical process like reionization.
} 
\maketitle
\section{Introduction}
\label{intro}
The relative paucity of dwarf galaxies is a well known challenge to the standard $\Lambda$CDM cosmological model.  The implication, within the standard paradigm, is that most collapsed dark matter halos are invisible.  They contain little cold gas and few stars.

There are a variety of plausible explanations.  For example there are the effects of reionization at redshift $6-7$ \cite{gne00,tho96}.  The thermal energy in re-heated gas prevented the gas from participating in collapse in shallow potentials.  Most halos collapsed after the epoch of reionization.  Those below a critical mass threshold of $\sim 2 \times 10^9 M_{\odot}$ would lack gas and, hence, stars.

There is the implication with this scenario that the fractional representation of visible dwarf galaxies will vary with environment.  Denser, generally more massive, environments collapsed earlier.  A larger (but still minor) fraction of dwarfs may already have been in place with gas before the onset of reionization in such places.  It was the contemplation of this possibility \cite{tul02} that motivated a program to search for dwarf galaxies in a wide range of circumstances.

New observational information comes predominantly from two sources.  One of these derives from wide field imaging of groups within the Local Supercluster with the Canada-France-Hawaii Telescope (CFHT).  Groups were studied with masses in the range $3 \times 10^{12} - 1 \times 10^{14}~M_{\odot}$ and with morphologies dominated alternatively by ellipticals and spirals.  Most of these groups lie at distances $\sim 25$~Mpc although a couple of the low mass groups lie closer.  The second source is provided by Hubble Space Telescope (HST) imaging.  Most galaxies within 4 Mpc have now been observed, allowing color--magnitude diagrams to be constructed that extend well below the tip of the red giant branch (TRGB) \cite{jac09}.  The luminosity of the TRGB provides an accurate measure of distance \cite{lee93,riz07}.  There is now detailed information about the clustering properties and luminosities of galaxies in the local volume \cite{kar05,tul06}.

\section{M81 Group}
\label{m81}

The M81 Group at 3.6 Mpc is amenable to observations that are both wide and deep.  To begin, an imaging survey of 65 sq. deg (525 by 525 kpc) was carried out at CFHT sensitive to $R \sim 25$, about 0.5 mag fainter than the anticipated magnitude of the brightest red giant branch stars at the distance of the group \cite{chi09}.  In addition to the 22 previously known M81 group members in the survey area, coincidently, 22 new candidates were identified.  These 22 candidates were subsequently observed with HST: 11 of 12 candidates considered likely members were confirmed, 3 of 6 uncertain candidates were confirmed, and 4 suspected artifacts were found to be such.  The 14 newly identified group members lie in the range $-12.5 < M_R < -6.9$.  The brightest of these has a UGC designation (UGC 5497) but had not been identified as nearby.  The faintest, that we call d0944+69, is the faintest known galaxy outside the Local Group.  It is considered that the survey is complete to $M_R = -9$.

\begin{figure}
\resizebox{1.05\hsize}{!}{\includegraphics*{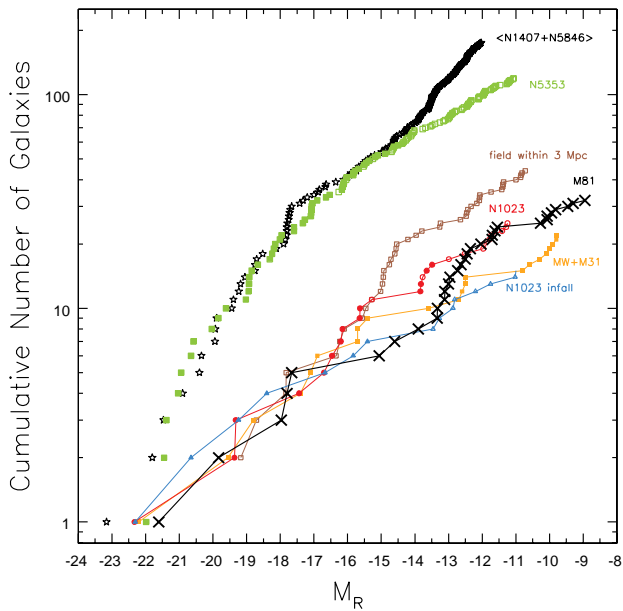}}
\caption{Cumulative luminosity functions for several nearby groups.  The luminosity function for the virial region around M81 (black crosses) extends to $M_R=-9$, fainter than the other samples.  Two relatively massive and dynamically evolved (overwhelmingly early typed) groups are combined to give the sample called $<$N1407+N5846$>$ (black stars).  The group labeled N5353 (green boxes) is relatively rich and contains both early and late components.  The other samples represent various low density and spiral dominated environments.}
\label{lf_m81plus}
\end{figure}

The cumulative luminosity function for the M81 Group is seen in Figure~\ref{lf_m81plus}.  It extends deeper than the other group luminosity functions assembled in the figure.  However the new and thought to be complete information does not substantially change what had previously been known.  The faint end Schechter function \cite{sch76} slope of $\alpha = -1.21\pm0.05$  is similar to what is found with other groups. 

\section{Inter Comparison of Several Groups}
\label{compare}

Fig.~1 includes cumulative luminosity functions for various other groups \cite{mah05,tre09,tre06,tul08}.  As a point of detail, the cumulative function is more useful than the familiar binned differential function if samples are small.  However faint end slope and bright end cutoff properties are less transparent.  On the other hand, the last point in a cumulative function directly gives the sample size at the faint limit of completion.

\begin{figure}
\resizebox{1.0\hsize}{!}{\includegraphics*{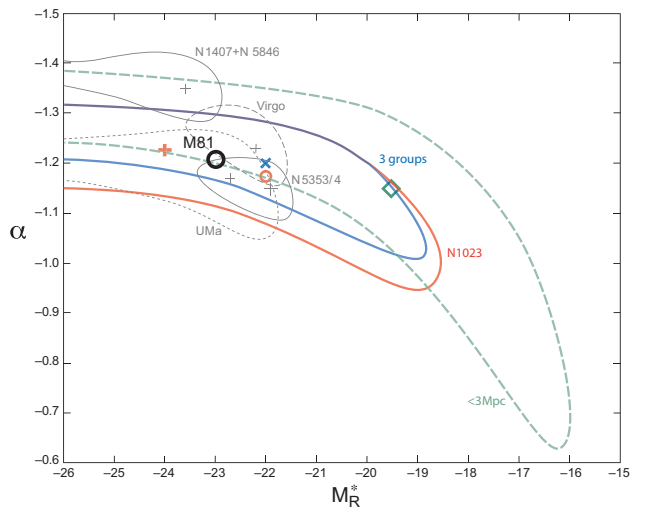}}
\caption{Comparison of the best fit Schechter function parameters, $M^{\star}$ and $\alpha$, for several nearby groups.  Contours identify 95\% probability limits.  The solid circle identifies the best fit for the M81 sample that is complete to $M_R = -9$.}
\label{Mstar_alpha}
\end{figure}

The Schechter function provides the standard characterization of the run of luminosities with magnitude.  There are two defining parameters besides the sample normalization:  the faint end power law slope, $\alpha$, where $\alpha=-1$ with equal numbers per magnitude bin and $\alpha=-2$ with a divergent luminosity in the faint end population; and the exponential bright end cutoff that becomes important above the magnitude $M^{\star}$.  The fits for the various samples are summarized in Figure~\ref{Mstar_alpha}.  The well known coupling between $M^{\star}$ and $\alpha$ is seen.  Steeper slopes are tolerated by brighter $M^{\star}$.  The contour levels are 95\% probability and it is apparent that $M^{\star}$ is usually unconstrained on the bright side (ie, a simple power law provides an adequate description) if a sample is small.

There are two points to take away from Figs. 1 and 2.  First, the range of $\alpha$ values is not large between groups and, yes, slopes are very significantly shallower than the $\alpha \sim -1.8$ anticipated for halo masses with hierarchical clustering in a $\Lambda$CDM universe.  Second, though subtle, the two more massive and much more dynamically evolved groups around NGC~1407 and NGC~5846 have slopes $\alpha \sim -1.35$ that are significantly steeper than the other groups (and the less well defined cutoffs $M_R^{\star} \sim -23.5$ are brighter).  

\begin{figure}
\resizebox{1.05\hsize}{!}{\includegraphics*{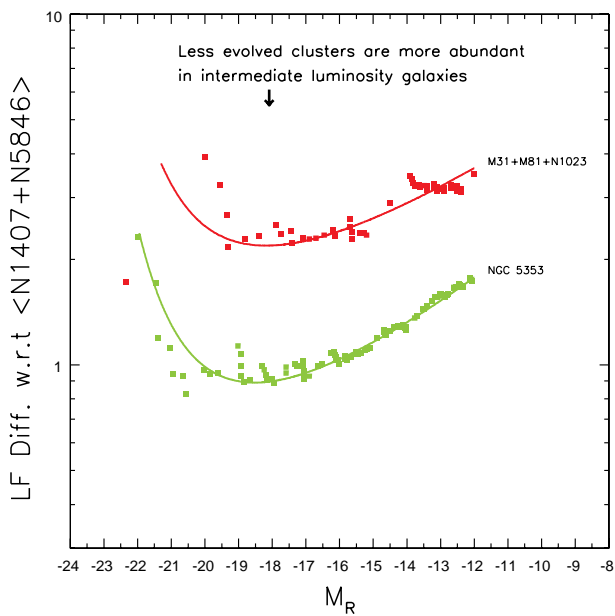}}
\caption{Differences in luminosity functions with the run of $M_R$.  The fiducial function is given by the combination of the NGC 1407 and NGC 5846 functions, two relatively massive ($8 \times 10^{13}~M_{\odot}$) groups composed almost exclusively of E/S0 and dE systems.  The NGC 5353 group (green crosses) and the combination of the M31, M81, and NGC 1023 groups (red boxes) are environments with prominent gas-rich components.  The minima at $M_R \sim -18$ is a consequence of the fact that the gas-rich environments contain relatively more intermediate luminosity systems and are relatively less well represented at high and low luminosities.}
\label{lfdif}
\end{figure}

Figure~\ref{lfdif} illustrates differences between group luminosity functions.  A reference luminosity function is formed by the combination of the NGC 1407 and NGC 5846 functions since they are the best populated and very similar to each other.  In the figure, one is given the differences between this composite reference function descriptive of dynamically evolved regions and two other functions drawn from relatively low density and unevolved regions.  The minima at $M_R \sim -18$ is a reflection of the fact that there are relatively more intermediate luminosity systems in the lesser evolved groups.  The difference in absolute numbers in comparison with the larger evolved samples is smallest at intermediate luminosities and increases going both brighter and fainter. 

\begin{figure}
\resizebox{1.01\hsize}{!}{\includegraphics*{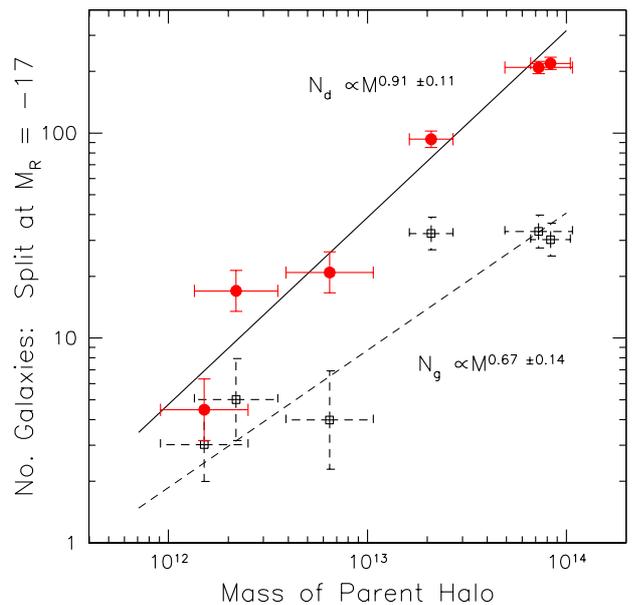}}
\caption{Numbers of dwarfs and giants as a function of the mass of the parent halo.  Halo masses are determined from the observed velocities of galaxies within the halo assuming an approximation to the virial condition.  Giant populations ($M_R < -17$) are identified in black and dwarf populations ($-17 < M_R < -11$) are in red.  The numbers of dwarfs is strongly correlated with halo mass.  Within $1 \sigma$ uncertainty, the number of dwarfs per unit halo mass is constant.}
\label{mass_numdw}
\end{figure}

Velocity information permits the determination of masses for each of the groups.  If one simply counts the numbers of giants and dwarfs in a group, with a break point at $M_R = -17$ (roughly the magnitude of the Small Magellanic Cloud), then a comparison of galaxy counts with the mass of the groups results in what is shown in Figure~\ref{mass_numdw}.  The correlation between group mass and number of giants is poor but the correlation between group mass and number of dwarfs is good.  It is of considerable interest that the slope of the dwarf correlation is unity within the errors.  There is a roughly constant number of dwarfs ($-17 < M_R < -11)$ per unit halo mass.

\section{Discussion}

All the group luminosity functions that have been studied have shallower faint end slopes than the theoretical $\Lambda$CDM mass function.  However the following significant variations must be reconciled.  In a comparison anchored by a constancy per unit halo mass in the number of dwarfs, evolved elliptical rich groups have relatively fewer intermediate luminosity galaxies and relatively more dominant principal galaxies than less evolved spiral-rich groups.  An evident possibility is that mergers are depleting the reservoir of sub-dominant galaxies in a halo while building the dominant galaxies.  Intermediate mass galaxies merge more quickly than lower mass systems through dynamical friction and because of lower velocities if there has been a drift toward energy equipartition from 2-body interactions \cite{bin87}.  Hence the observed qualitative differences in luminosity functions can be understood. 

The constancy of dwarf numbers per unit halo mass was a bit of a surprise.  In any event, the depletion of smaller systems with time in a halo must be inevitable (modulo replenishment through infall).  Group halos that collapsed earlier presumably have experienced greater depletion.  It follows that the constancy of dwarf galaxies per unit mass seen in the groups that were studied is not reflective of earlier conditions.  The groups that formed earliest probably had a higher fraction of star-filled dwarfs per unit mass at inception.

The ultimate question is whether the universally rather flat faint end luminosity function slope of $\alpha \sim -1.2$ to $-1.35$ poses a grave challenge to the standard paradigm of hierarchical clustering.  The point has been made by Tolstoy and Grebel at this conference that although every well studied nearby dwarf has an ancient population, typically only 20\% of the stars were in place near in time to the epoch of reionization.  The pronounced trend of lower metallicities in less luminous dwarfs \cite{mat98} implies a circulation of the gas reservoir that births later generations; involving expulsion of metal enriched gas and replenishment diluted by pristine intergalactic material.  Most stars in the dwarfs that have been adequately studied are created by this process at low redshift. If pristine gas continues to accumulate in low mass halos at low redshifts, it is not clear that the reionization mechanism suffices to explain the lack of visible dwarfs.  In the summary to the meeting, Simon White  wrote in bold letters on a powerpoint slide and emphasized verbally that ``$\Lambda$CDM must be wrong''.

\bigskip\noindent
Kristin Chiboucas, Brad Jacobs, Igor Karachentsev, Andi Mahdavi, Luca Rizzi, and Neil Trentham have participated in the observations discussed in this paper.  Those observations involved use of the Canada-France-Hawaii and Subaru telescopes at Mauna Kea Observatory and Hubble Space Telescope.  Financial support has been provided by Hubble Space Telescope awards AR-11285 and GO-11584 and from the National Science Foundation award AST-0908846.

\end{document}